\documentclass{emulateapj}
\usepackage{natbib}
\usepackage{apjfonts}
\slugcomment{Accepted June 8, 2011}

\shorttitle{The Distance of Cygnus X-1}
\shortauthors{Xiang et al.}

\def\cygx1{Cygnus~X-1\,} 
\def\chandra{{\it Chandra}\,}

\begin{document}

\title{Using the X-ray Dust Scattering Halo of Cygnus X-1 \\ 
to determine distance and dust distributions}

\author{Jingen Xiang\altaffilmark{1}, Julia C. Lee\altaffilmark{1}}
\affil{\altaffilmark{1}Harvard University Department of Astronomy
(a part of the Harvard-Smithsonian Center for Astrophysics), 
60 Garden Street, Cambridge, MA 02138}
\email{jxiang@cfa.harvard.edu; jclee@cfa.harvard.edu}
\and
\author{Michael A. Nowak\altaffilmark{2}, J\"orn Wilms \altaffilmark{3}}
\affil{\altaffilmark{2}Massachusetts Institute of Technology, Kavli Center, 
77 Massachusetts Ave.\ NE80, Cambridge, MA 02139}
\affil{\altaffilmark{3}Dr.\ Karl-Remeis-Observatory and Erlangen
  Centre for Astroparticle Physics, University of Erlangen-Nuremberg,
  Sternwartstr.~7, 96049 Bamberg, Germany}

\begin{abstract}
  We present a detailed study of the X-ray dust scattering halo of the
  black hole candidate \cygx1 based on two \chandra\ HETGS
  observations. Using 18 different dust models, including one modified
  by us (dubbed XLNW), we probe the interstellar medium between us and this
  source. A consistent description of the cloud properties along the
  line of sight that describes at the same time the halo radial
  profile, the halo lightcurves, and the column density from source
  spectroscopy is best achieved with a small subset of these models.
  Combining the studies of the halo radial profile and the halo
  lightcurves, we favor a geometric distance to \cygx1 of
  $d=1.81\pm{0.09}$\,kpc.  Our study also shows that there is a dense
  cloud, which contributes $\sim$50\% of the dust grains along the
  line of sight to \cygx1, located at $\sim1.6$ kpc from us. The
  remainder of the dust along the line of sight is close to the black
  hole binary.
\end{abstract}


\section{Introduction}
It has long been known that X-ray sources with high column densities
are associated with dust ``scattering halos'', which are formed by the
scattering of the sources's X-rays from a foreground dust containing
cloud in the interstellar medium (ISM). First predicted by
\citet{overbeck65} and then expanded upon by \citet{trumper73},
scattering halos were first observationally confirmed by
\citet{rolf83} using an \textit{Einstein} X-ray Observatory
observation of the X-ray binary GX~339$-$4.  Such halos were later
studied with virtually all imaging X-ray instruments
\citep{predehl95,predehl96,predehl00,costantini05}.

The properties of a dust scattering halo, i.e., the halo radial
profile and the delay between source lightcurves and halo lightcurves,
depend upon the composition, size distribution, and spatial
distribution of the intervening, scattering dust grains. Therefore
scattering halos have been widely used to probe the porosity of grains
\citep{mathis95,smith98,smith02}, grain composition
\citep{costantini05}, and the spatial distribution of the dust along
the line-of-sight \citep[LOS;][]{xiang05, xiang07, ling09}.
Comprehensive studies of dust halos include the systematic
\textit{ROSAT} studies of \citet{predehl95} --- yielding relationships
between the equivalent hydrogen column, $N_{\rm H}$, and the dust
grain size distribution in the interstellar medium (ISM) --- and the
\chandra\ ACIS-S/HETGS studies of \citet{xiang05}, for the
determination of dust spatial distributions. See also \citet{xiang07}
where non-uniform distributions were explored in the context of the
structure of the Milky Way.

We note, however, that many of these results also depend upon the
properties of the dust assumed in the modeling. Comparing
reddening-based hydrogen column densities $N_{\rm H}$ with dust models
in the case of 4U\,1724$-$307 and X Persei, \citet{valencic08} and
\citet{valencic09} showed that only some of the models summarized by
\citet[][hereafter ZDA]{ZDA} are consistent with the measurements,
while the commonly-used models of \citet[][hereafter MRN]{MRN} and
\citet[][hereafter WD01]{WD01} significantly underestimated the
measured $N_{\rm H}$.

The \chandra\ Advanced CCD Imaging Spectrometer (ACIS;
\citealt{garmire:03a}), with its high angular and energy resolution,
is perhaps the best instrument to observe spatially resolved X-ray
dust halos. Furthermore, via timing analysis of the halo these data
can be used to determine the distance to a bright, variable source.
One such distance measurement was presented by \citet{predehl00} for
an observation of Cygnus X-3, using the method proposed by
\citet{trumper73}. Since then, this method has been used to determine
the distances to several X-ray sources, including 4U~1624$-$490
\citep{xiang07} and Cen X-3 \citep{thompson09}. In the case of
4U~1624$-$490, the distance estimate to the object was improved
from about 30\% uncertainty to 15\%.

\cygx1 is a well known high-mass X-ray binary (HMXB) that includes a
blue supergiant star and a black hole candidate. Although it was
discovered in 1964, its distance has been uncertain. Distance
estimates ranged from 2.5$\pm$0.4~kpc \citep[derived from optical
  extinction measurements;][]{margon73} to $1.4_{-0.4}^{+0.9}$~kpc
\citep[based on multiple-epoch phase referenced VLBI
  observations;][]{lestrade99}. Most workers seem to have resorted to
$2.0\pm0.1$~kpc, which, e.g., has been used to estimate the position
of the scattering dust \citep{ling09}.  Very recent estimates, based
upon parallax measurements of the radio jet, place \cygx1 at
$1.86\pm0.12$\,kpc (Reid et al., 2011, submitted).  In
Section~\ref{sec:distance} we present distance determinations
independent of all of these estimates, and demonstrate that the halo
method yields a distance estimate of comparable quality to the
parallax method.

\begin{figure*}
\plotone{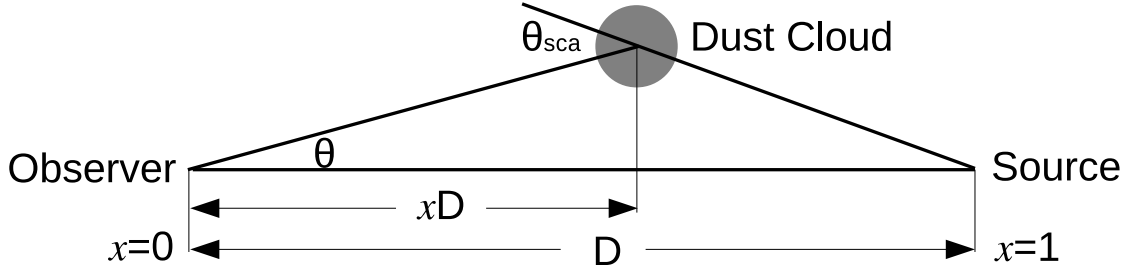}
\caption{Geometry of the X-ray-scattering process for single
  scattering.\label{fig:sca}}
\end{figure*}

In this paper, we first give a brief theoretical description of dust
halos in Section \ref{sec:theory}.  Then, using two
\chandra\ ACIS-S/High Energy Transmission Gratings Spectrometer
(HETGS; \citealt{canizares2005}) observations (Section
\ref{sec:analysis}, Section \ref{sec:spatial}), we present a study of
the composition, density, and spatial distribution of the dust grains
along the line of sight (LOS) based upon the halo radial profile
(Section \ref{sec:spatial}).  We then give a detailed study for the
distance determination based on halo timing analysis and show that
this method is very reliable (Section \ref{sec:distance}).  We
summarize our conclusions in Section \ref{sec:discuss}.

\section{Theoretical and Historical Background} 
\label{sec:theory}

The theory governing the observed halo surface brightness has been
well explored by many authors \citep[e.g.,][and references
therein]{overbeck65, mauche84,mathis91, predehl96, smith02} and the
calculation of the time delay of a scattered photon with respect to an
unscattered one also has been discussed extensively \citep{trumper73}.
In the following, we give a brief summary of the main points of this
theory required for our analyses. Readers desiring a more rigorous
explanation are referred to the relevant papers cited above. See also
\cite{xiang07} for details (including a flow chart) of the analysis
sequence.

\subsection{Halo Surface Brightness}
\label{sec:halo_s}

X-ray scattering halos are formed when X-rays emitted by a background
source are scattered by dust in the intervening ISM. As discussed by
\citet{mathis91} and \citet{smith02}, for a single scattering
(Fig.~\ref{fig:sca}) at an observed angle $\theta$, the observed
first-order scattering halo surface brightness for photons with energy
$E$ can be described by
\begin{eqnarray}\label{surface_I}
I_\mathrm{sca}^{(1)}(\theta, E) = F_\mathrm{X}(E)
N_\mathrm{H}\int_{a_\mathrm{min}}^{a_\mathrm{max}}
 n(a) \, \int_{0}^{1}{f(x)(1-x)^{-2}}  \nonumber \\ \times  
S\left(a,E,{\theta_{\mathrm{sca}}}\right) \,da\,dx ~~,
\end{eqnarray}
where $F_\mathrm{X}(E)$ is the total observed X-ray photon flux (in
units of $\rm photons\ cm^{-2}\ s^{-1}$) at energy $E$, the
relative distance $x=d/D$ is the ratio of the distance from scattering
grain to the observer ($d$) and the distance from the source to the
observer ($D$), $N_{\rm H}$ is the equivalent hydrogen column density
between the observer and the X-ray source, and $f(x)$ is the ratio of
the local hydrogen density at $xD$ to the average hydrogen density
along the LOS (Fig. \ref{fig:sca}). 
The grain properties are defined via the size distribution of the dust
grains, $n(a)$, and the energy-dependent differential cross section at
a scattering angle of $\theta_{\mathrm{sca}}$ for a spherical particle of
radius $a$, $S(a, E, \theta_{\mathrm{sca}})$, which in the Rayleigh-Gans
approximation is given by \citep{mathis91}
\begin{eqnarray}
S(a,E,\theta_\mathrm{sca})=\frac{d\sigma_\mathrm{sca}(a, E,
  \theta_\mathrm{sca})}{d\Omega} 
= 
c_{1}
\left(\frac{2Z}{M}\right)^{2} 
\nonumber \\  
\left(\frac{\rho}{3\,\mathrm{g}\,\mathrm{cm}^{-3}}\right)^2 
\left(\frac{a}{\mu m} \right)^{6}
\left[\frac{F(E)}{Z}\right]^{2}\exp({-K^{2}\theta_\mathrm{sca}^{2}}), 
\end{eqnarray}
where $c_1 = 1.1 \times 10^{-12}\,{\rm cm^2/sr}$, $K=0.4575(E/\mathrm{kev})^2
(a/\mu\mathrm{m})^2$, $Z$ is the mean atomic charge, $M$ is the
molecular weight (measured in atomic mass units), $\rho$ is the mass
density and $F(E)$ is the atomic scattering factor \citep[taken, e.g.,
from][]{henke81}. Thus a photon of wavelength $\lambda$ will be
scattered into a typical angle
\begin{equation}
  \theta=\frac{\lambda}{\pi\ a} \propto E^{-1}.
\end{equation}
For small angles (several arcmin), $\theta \sim(1-x)
\theta_\mathrm{sca}$.

\citet{mathis91} have shown that for scattering optical depths greater
than 1, doubly scattered radiation may dominate the multiple
scattering at distances of several arcmin. For the scattering halo
discussed here, however, even at energies as low as 1\,keV the optical
depth, $\tau<1$. Multiple scattering therefore can be neglected for
the photon energies considered in this work.

\begin{figure}
\plotone{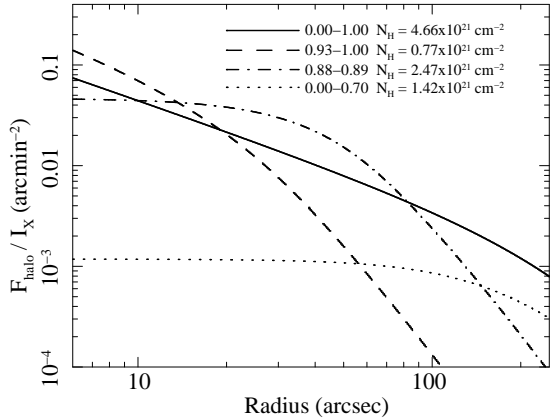}
\caption{Halo radial profiles for the different spatial distributions
  of the dust grains based on the dust model of ZDA BARE-GR-S. (See
  \S\ref{sec:dustmodels} for a discussion of this model.) The hydrogen
  column densities are based on the halo fits discussed in
  \S\ref{sec:distance}, and the indicated ranges are the relative
  distance between observer, at 0, and source, at 1. The specific
  model shown in fact corresponds to the first entry in
  Table~\protect{\ref{tab:nh}}.\label{fig:spatial}}
\end{figure}

Eq.~\ref{surface_I} shows that the halo surface brightness, as well
the time delay of the scattering photon (see \S\ref{sec:time} below),
tightly depends upon the spatial distribution of the dust grains and
on the size distribution of the grains. As an example,
Fig.~\ref{fig:spatial} shows halo radial profiles for different
spatial distributions. Halos due to dust clouds close to a source are
sharp and narrow, while halos due to clouds closer to the observer are
flat and wide.

\subsection{Time Delay of the Scattering Photon}
\label{sec:time}

From the geometry shown in Fig.~\ref{fig:sca}, the total distance
traveled by a photon scattered at $xD$ is
\begin{equation}
D^{'} = \frac{xD}{\sin(\theta)} + \sqrt{(D-xD)^2 + (xD\ \tan\theta)^2} ~~.
\end{equation}
This photon will travel a longer distance than the unscattered
one. For small angles, $\theta\ll 1$ and
$\sin(\theta)\sim\tan(\theta)\sim\theta$. In this case, the delay time
is
\begin{equation}
\label{eq:delay}
   dt = \frac{D^{'} - D}{c} = 1.21\,\mathrm{sec} \left ( \frac{D}{1\,\mathrm{kpc}} \right)
\left(\frac{\theta}{1^{''}}\right)^2 \frac{x}{1-x}
\end{equation}
The delay time therefore increases dramatically at large angles and
easily becomes greater than the time scales of the intrinsic intensity
variations. For this reason, if one wants to determine the distance
from time delay measurements by inverting Eq.~\ref{eq:delay}, the
variability of the halo at small scattering angles has to be used. In
addition, as shown in Fig.~\ref{eq:delay} the delay time tightly
depends on the position of the scattering medium. For this reason, the
distance and the spatial distribution of the dust have to be
determined together.

\subsection{Dust Grain Models}\label{sec:dustmodels}
As discussed above, dust composition and grain size distribution
affect the overall determination of the halo properties. In this
section, we will briefly describe the most commonly used models for
dust in the ISM. These include the two most commonly used grain models
of MRN and WD01, as well as a sequence of 15 different kinds of grain
models recently constructed by \citet{ZDA}.

The classical grain model is that of MRN, which assumes both graphite
and silicate grains with a power law distribution. Based on the
observed interstellar extinction over the wavelength range $0.11\,\mu
\mathrm{m} < \lambda < 1\,\mu\mathrm{m}$, the size distribution is
described by $n(a)=K (a/1\,\mu\mathrm{m})^{-3.5}$, where $K =
10^{-15.24}$ for graphite and $K = 10^{-15.21}$ for silicate. Both
size distributions are sharply cut off outside of $0.005\,\mu\mathrm{m} <
a < 0.25\,\mu\mathrm{m}$.

\begin{figure}
\plotone{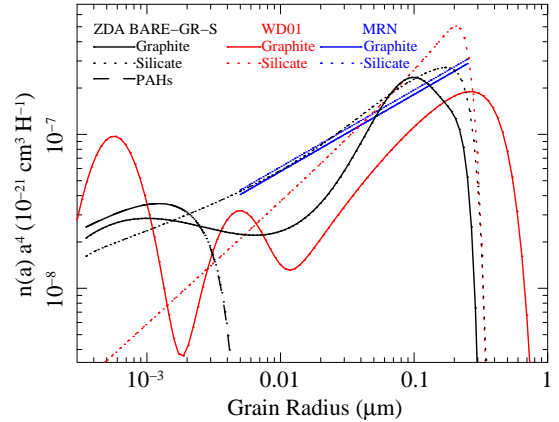}
\caption{Example grain-size distributions for the MRN, WD01, and the
  ZDA BARE-GR-S dust models.\label{fig:dust}}
\end{figure}

\citet{WD01} constructed grain models for different regions of the
Milky Way, LMC, and SMC. In contrast to the MRN model, the WD01 model
includes sufficiently small carbonaceous grains (including polycyclic
aromatic hydrocarbon molecules, PAHs) to account for the observed
infrared and microwave emission from the diffuse ISM. The size
distributions of both carbonaceous and silicate grains are not simple
power laws (see Fig.~\ref{fig:dust}). 

Unlike MRN and WD01 models, which based elemental abundance constraints upon
only one interstellar medium abundance, \citet{ZDA} include solar (S),
F- and G-star (FG), and B-star (B) abundances when they derive their
interstellar dust models. The ZDA model is derived from simultaneous
fits to the far-ultraviolet to near-infrared extinction and assumes
five different dust constituents: (1) PAHs; (2) graphite; (3)
amorphous carbon; (4) silicates in the form of olivines ($\rm MgFeSiO_4$); 
and (5) composite particles containing different proportions of silicates,
organic refractory material ($\rm C_{25}\rm H_{25}\rm O_5\rm N$), water ice
($\rm H_2O$), and voids. ZDA consider two groups of models, one group
that include only PAHs and bare grains and another group containing
further composite particles. These groups are called BARE and COMP,
respectively. The BARE and COMP models are further subdivided
according to the properties of carbon in the different models. ZDA
distinguish between graphite (GR), amorphous carbon (AC), and no
carbon (NC). Taking into account the different abundances (designated
previously as -S, -FG and -B), a total of 15 different dust grain
models are obtained. The designation of each model derives from the
abbreviations listed above. For example, the model consisting of bare
silicate, PAHs and graphite and derived by assuming  F- and G-star
abundances will be called BARE-GR-FG.

In order to compare these three kinds of models with each other,
Fig.~\ref{fig:dust} shows their size distributions. In order to avoid
unnecessary clutter in the figure, for the ZDA models we only plot the
size distribution of BARE-GR-S. This latter model yields the proper
reddenings along the LOS to the low-mass binary 4U~1724$-$307
\citep{valencic09}. We also list the key parameters, e.g.,
composition, size range and abundance of dust grain, in Table
\ref{tab:model}. The parameters of models 1 to 15 are taken from
\citep{ZDA}, MRN from \citep{MRN}, and WD01 from \citep{WD01}
respectively. The model of ``XLNW'' is a modified form of ZDA
BARE-GR-S (listed as model 1 in the tables; see \S\ref{sec:spatial}
for a detailed discussion).

\begin{deluxetable*}{rlccccccc}
\tabletypesize{\scriptsize}
\tablecaption{Summary of dust models.}
\tablecolumns{9}
\tablehead{\colhead{Model$^a$} & 
\colhead{Model$^b$} & 
\colhead{PAH$^c$} & 
\colhead{Graphite$^d$} &
\colhead{ACH2$^e$} & 
\colhead{Olivine$^f$} & 
\colhead{Composite$^g$} &
\colhead{Iron$^h$} & 
\colhead{Reference$^i$}\\
\colhead{No} & 
\colhead{Name} &
\colhead{Size Range} & 
\colhead{Size Range} & 
\colhead{Size Range} & 
\colhead{Size Range} &
\colhead{Size Range} &
\colhead{} &
\colhead{} \\ 
\colhead{} & 
\colhead{} & 
\colhead{Abundance} &
\colhead{Abundance} &
\colhead{Abundance} & 
\colhead{Abundance} & 
\colhead{Abundance} &
\colhead{Abundance} &
\colhead{}}
\startdata
1  &  BARE-GR-S  &  0.35--5.0   &  0.35--330  & \nodata  &  0.35--370 & \nodata & \nodata & 1  \\
    &                      &  33.0  &  212.9 & \nodata  & 33.3 &\nodata & 33.3 \\
2  & BARE-GR-FG  & 0.35--5.0  & 0.35--300 & \nodata & 0.35--340  & \nodata & \nodata  & 1  \\
    &                      &  35.2 & 212.3 & \nodata  & 33.1 & \nodata & 33.1 \\
3  & BARE-GR-B  & 0.35--3.5 & 0.35--320 & \nodata & 0.35--320 & \nodata & \nodata   & 1  \\
    &                      & 33.3	& 221.1 & \nodata & 28.5 & \nodata  & 28.5 \\
4  & COMP-GR-S & 0.35--5.5 & 0.35--500 & \nodata & 0.35--440 & 20--900 & \nodata  & 1  \\
    &                      & 33.5 & 109.2 &  \nodata & 25.0 & 8.0 & 33.0\\ 
5  & COMP-GR-FG & 0.35--5.5 & 0.35--390 & \nodata & 0.35--390 & 20--750 & \nodata   & 1 \\
    &                      & 35.8  & 133.3 & \nodata & 26.1 &  6.3 & 32.4\\
6  & COMP-GR-B & 0.35--5.5 & 0.35--520 & \nodata & 0.35--330 & 20--450 & \nodata   & 1 \\
    &                     & 33.7  &  133.0 & \nodata & 20.0 & 7.8 & 27.8 \\
7  & BARE-AC-S & 0.35--3.7 & \nodata & 20--260 & 3.5--370 & \nodata & \nodata  & 1 \\
    &                     & 51.4  &  \nodata & 213.6  & 33.5 & \nodata & 33.5\\
8  & BARE-AC-FG & 0.35--3.6 & \nodata & 20--280 & 3.5--370 & \nodata & \nodata  & 1 \\
    &                    & 52.4  &  \nodata & 212.7  & 34.2  & \nodata & 34.2\\
9  & BARE-AC-B & 0.35--3.7 & \nodata & 20--250 & 3.5--330 & \nodata & \nodata  & 1 \\
    &                    & 52.2  & \nodata & 223.1 & 28.7  & \nodata & 28.7\\
10 & COMP-AC-S & 0.35--3.8 & \nodata & 20--250 & 0.35--400 & 20--910 & \nodata  & 1 \\
    &                    &  50.6 & \nodata & 75.2 & 23.7 & 9.6 & 33.5\\
11 & COMP-AC-FG & 0.35--3.5 & \nodata & 20--250 & 0.35--400 & 20--660 & \nodata  & 1 \\
    &                    &  51.7 & \nodata & 81.2 & 24.5 & 9.2 & 33.7\\
12 & COMP-AC-B & 0.35--3.9 & \nodata & 22--210 & 0.35--250 & 20--700 & \nodata  & 1 \\
    &                    &  51.5 & \nodata & 28.1 & 14.3 & 13.6 & 27.9\\
13 & COMP-NC-S & 0.35--3.6 & \nodata & \nodata & 0.35--340 & 20--800 & \nodata  & 1 \\
    &                    &  50.0 & \nodata & \nodata & 18.7 & 14.7 & 33.4\\
14 & COMP-NC-FG & 0.35--3.5 & \nodata & \nodata & 0.35--360 & 19--850 & \nodata  & 1 \\
    &                    &  51.0 & \nodata & \nodata & 19.1 & 14.8 & 33.9\\
15 & COMP-NC-B & 0.35--3.8 & \nodata & \nodata & 0.35--180 & 20--800 & \nodata  & 1 \\
    &                    & 51.5 & \nodata & \nodata & 12.7 & 15.4 & 28.1\\
16 & MRN           & \nodata &  5--250 & \nodata & 5--250 & \nodata & \nodata  & 2 \\
    &                    & \nodata &  270.0  & \nodata & 33.0 & \nodata & 33.0\\
17 & WD01         & \nodata & 0.35--1000 & \nodata & 0.35--400 & \nodata & \nodata  & 3 \\
    &                    & \nodata & 231.0 & \nodata & 36.3 & \nodata & 36.3\\ 
\hline
     &                 &  PAHs   & Graphite   &  ACH2  & Enstatite & Fe metal  \\
\hline 
18 & XLNW       &  0.35--5.0 & 0.35--330 & \nodata & 0.35--370 & 0.35--370  & \nodata  & 4  \\
     &                 &  33.0   & 212.3  & \nodata & 33.3 & 33.3 & 33.3\\
 \enddata
\tablenotetext{a,b}{Model number and model name}
\tablenotetext{c--g}{The unit of the size range is $10^{-3}\ \mu$m (nm)
and the abundance is the molecular number of composition per hydrogen
in ppm ($10^{-6}$). The abundance of composites is based on
the abundance of olivine residing in the composite.}
\tablenotetext{f} {The silicate in the ZDA, MRN and WD01
models is explicitly shown as olivine (MgFeSiO$_4$) in order to
discriminate it from other silicates, e.g., enstatite (MgSiO$_3$) in
the XLNW model.}  
\tablenotetext{h} {The iron abundance equals to the sum of abundances of all iron compounds, e.g., olivine, composite and iron metal. The unit is also ppm ($10^{-6}$).}
\tablenotetext{i}{References for dust model parameters:
1. \citet{ZDA}, 2. \citet{MRN}, 3. \citet{WD01} and
4. this paper.}  \label{tab:model}
\end{deluxetable*}

Based on a detailed analysis of observed XAFS (X-ray Absorption Fine
Structure) in Cygnus X-1 spectra of the Fe L band \citep{lee:09b,lee11} using
the techniques described in \citet[][see also
  \citealt{lee:10a}]{lee:09a}, we introduce an additional model, which
we dub XLNW, where the major Fe-containing grain compound olivine
($\rm FeMgSiO_4$) in the ZDA BARE-GR-S model is replaced with an iron
grain consisting of a metallic iron core surrounded by troilite (FeS),
and enstatite ($\rm MgSiO_3$).  In generating this new model, for each
compound, the normalization coefficient ($A$) is adjusted to meet the
criteria of the following dust mass equations (in different
representations on both the left and right side)
\begin{equation}
\frac{A_Z ~ M}{N_{\rm A}} =  {{{4} \over {3}}\pi \rho  
\int_{a_{\rm min}}^{a_{\rm max}}{\rm d}a\, n(a)a^3},
\end{equation}
where for the grains replacing olivine, $A_{\rm Z} = A_{\rm Mg} = A_{\rm Si} = A_{\rm Fe}$ 
is the abundance of the dust compound, $M$ is the molecular 
weight in atomic mass units (AMU), and $N_{\rm A}=6.02\times 10^{23}$ is 
Avogadro's constant. As described above, $\rho$, $a$, $a_{\rm min}$, 
$a_{\rm max}$ and $n(a)$ are, respectively, mass density, radius, 
minimal radius, maximal radius and size distribution of the dust grains. 
(The normalization coefficient, $A$, is a parameter contained within $n(a)$.)
For the iron compound, i.e., iron metal + troilite ($\rm FeS_{0.6}$), 
$\rho = 5.2~{\rm g \, cm^{-3}}$, $A_Z=33.3\times10^{-6}$ and $M=75.2$~AMU while
$\rho = 3.2~{\rm g \, cm^{-3}}$, $A_Z=33.3\times10^{-6}$ and $M=100$~AMU for
enstatite ($\rm MgSiO_3$). Values for $\rho$ are taken from a mineralogy
database\footnote{http://webmineral.com/}; where the compound is an
admixture, $\rho$ is taken to be the weighted average of the minerals
which make up the compound. 

\begin{figure}
\plotone{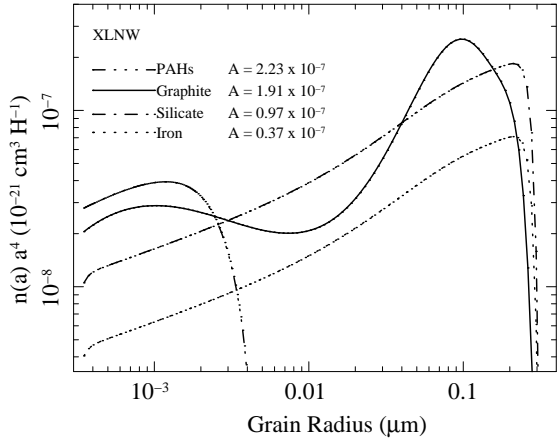}
\caption{The particle size distributions of our XLNW dust model, which is a
  modified form of the ZDA BARE-GR-S model.\label{fig:mod}}
\end{figure}

As in the ADA and MRN models, all Fe and Si are
assumed to reside in the dust, as borne out also by our XAFS fitting.
The particle size distributions of iron dust and enstatite are the
same as those of silicate in ZDA BARE-GR-S. The normalization of the
size distribution, however, has been slightly changed by adjusting the
amount of dust appropriate for the assumed depletion. The size
distributions and the normalization coefficient ($A$) of the XLNW dust
model are shown in Fig.~\ref{fig:mod}.

\section{Observation and Data Analysis}
\label{sec:analysis}
We now will use two \chandra-HETGS observations to study the X-ray
halo of \cygx1 using the methods discussed above. The first
observation occurred on 1999 October 19 (ObsID 107) and lasted about
15\,ks.  During this time the source was transiting between its hard,
power-law dominated state, and the more thermal soft state (see
\citealt{schulz02} for a study of the source behavior). As is
characteristic for black holes in this transitional state, the overall
source variability was low. This property allows us to use these data
to measure a single non-variable radial halo profile which we use to
determine the spatial distribution of the dust grains. The second
observation was performed on 2004 April 19 (ObsID 3814). This later
observation lasted about 50\,ks and was designed by us to be performed
during the upper conjunction of the black hole
\citep{manfred09}. During this orbital phase the X-ray lightcurve is
strongly variable due to frequent photoabsorption dips as the line of
sight to the black hole crosses through the clumpy focused stellar
wind of the companion. Due to dramatic variability, this observation is
well-suited to determine the distance to \cygx1.

We create two datasets, as described below, to perform these
analyses. One is a steady spatial/spectral halo radial profile that is
used for determining the composition of the halo. This dataset is
primarily comprised of ObsId 107, however, we describe a procedure to
replace the inner ($<$20~arcsec) data in order to obtain a more
accurate profile.  The second dataset consists of variable halo
lightcurves, as a function of radius, comprised solely of data from
ObsID 3814.  The latter dataset is used for the distance
determination.

In order to form the datasets, we follow the same procedure described
in \citet{xiang07} to extract the HETGS spectra, halo radial profile,
and halo lightcurve. (See that paper for a detailed analysis flow
chart.) CIAO 4.2 with CALDB 4.2 was used. A theoretical point spread
function (PSF) created with the Chandra Ray Tracer (ChaRT;
\citealt{carter:2003}) is used for small angles ($<$60~arcsec). ChaRT
has been shown to represent properly the PSF behavior;
\citep{xiang07}.  The PSF obtained from ChaRT simulations, however, is
underestimated at large off-axis angles ($\gtrsim60''$). We therefore
use the bright and halo-less point source Her~X-1 to perform an
\textsl{ab initio} determination of the PSF at large angles. Her~X-1
was observed on 2002 July 02 (ObsID 3662) with \chandra ACIS-I. The
parameters of these three observations are listed in
Table~\ref{tab:observation}.

\begin{deluxetable*}{lcllllll}
\tabletypesize{\scriptsize}
\tablecaption{Parameters of three \chandra\ observations}
\tablecolumns{8}
\tablehead{
\colhead{Source} &
\colhead{Observation} &
\colhead{State} &
\colhead{Instrument} &
\colhead{Start Time} &
\colhead{End Time} &
\colhead{Exposure} &
\colhead{Usage$^a$}\\
\colhead{} &
\colhead{ID} &
\colhead{} &
\colhead{} &
\colhead{} &
\colhead{} &
\colhead{(ks)} &
\colhead{}}
\startdata
Cyg X-1   & 107  & Transition & AICS-S/HETG  & 1999-10-19 19:17:37   &
1999-10-19 23:52:10   & 11.4  & RP\\
Cyg X-1   & 3814 & Low/Hard   & AICS-S/HETG  &  2003-04-19 16:47:31 &
2003-04-20 06:41:42 &  47.2 & RP + LC\\
Her X-1   & 3663 & High/Soft & AICS-S/I  & 2002-07-01 23:37:38  &
2002-07-02 14:03:05 & 49.6 & PSF\\
\enddata
\tablenotetext{a}{ ``RP" means that the data are used to generate
the``Halo Radial Profile",``LC" for "Halo Lightcurves" and ``PSF" for``Point Spread Function", respectively.}
\label{tab:observation}
\end{deluxetable*}

\begin{figure}
\plotone{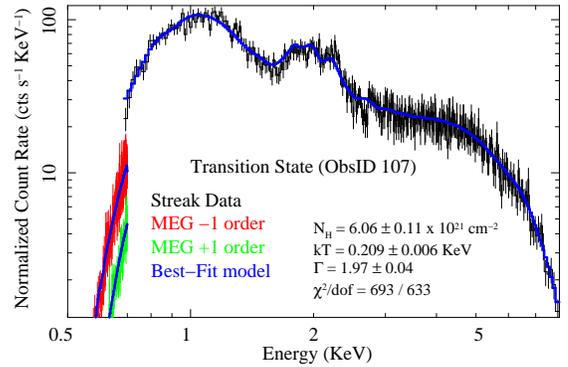}
\caption{The streak spectrum and MEG first order spectra overplotted with
the best fit model.\label{fig:spectra}}
\end{figure}

Both the halo and the PSF represent functions of radius and
energy. The PSF specifically is the ratio between the surface
brightness, absent a halo, and the point source spectrum.  In order to
determine the surface brightness associated with the combined halo and
PSF, we need an accurate estimate of the source spectrum for both Cyg
X-1 observations discussed here.  In each of these cases, the spectra
extracted from the zeroth-order suffer heavy pileup over the entire
energy band and also suffer pileup in the first order gratings spectra
above 0.7~keV.  We therefore follow the method used by \cite{smith02}
and \cite{valencic09} to extract the spectrum from the readout streak,
which is completely pileup free.  We then use the simple model -- a
disk blackbody plus powerlaw coupled with cold gas absorption -- that
has been used to fit these two observations
\citep{schulz02,manfred09}. Below 0.7\,keV, we use the MEG$\pm$1
spectra, which are pileup free.  The flux of \cygx1 is determined from
the fit to these joint MEG$\pm1$ spectra and the readout streak
spectra. The transitional state spectra, overplotted with the best fit
model, are shown in Fig. \ref{fig:spectra}.  (The spectral flux during
the hard state is also derived with the same methods.)  It should be
noted that the $N_{\rm H}$ fit from these spectra contains gas local
to the \cygx1 system and therefore is expected to result in $N_{\rm
  H}$ values higher than those that we expect to obtain from modeling
the halo radial profile.

\begin{figure}
\plotone{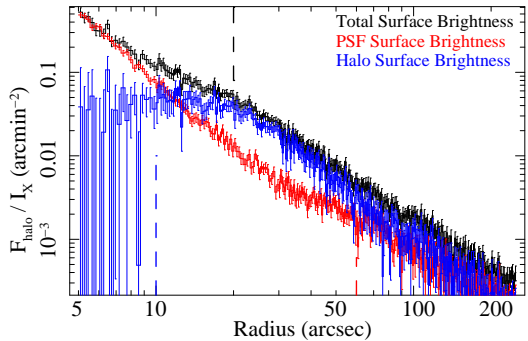}
\caption{Comparison of the radial profiles between the halo and the
  PSF (i.e., the surface brightness attributable solely to the point
  source) in the 2.0--2.2\,keV energy band. The total surface
  brightness (black) was created from a combination of ObsID 3814
  ($<$20~arcsec) and ObsID 107 ($>$20~arcsec).~--~A vertical black
  dashed line delineates these two regions. The PSF profile was
  created from ChaRT simulations ($<$60~arcsec) and from a
  non-variable Her~X-1 observation ($>60$~arcsec). Data from less than
  $10$~arcsec (marked by vertical blue dashed line) is not considered
  in this energy band.\label{fig:extraction}}
\end{figure}

As discussed above, ObsID 107 was used to generate the non-variable
halo radial profiles.  Due to the brightness of Cygnus X-1 during the
transitional observation (ObsID 107), however, these data suffer
severe pileup at small angles. In order to investigate clouds very
near the source, halos at small angles are needed.  We therefore
excise the core of the halo from ObsID 107 and substitute data from
the hard state observation, where due to its low flux, pileup free
halos can be extended to as low as $7''$ radius. In this work, the
total surface brightness (see Fig.~\ref{fig:extraction}) at greater
than 20~arcsec radius is extracted from the transitional state
observation (ObsID 107), while the surface brightness at less than
20~arcsec radius is from the hard state (ObsID 3814).  Even though
\cygx1 is variable, this approach is justified since at small angles
the time delay between the source and the halo is negligible, and we
therefore can compare the time averaged source flux to the time
averaged radial profiles.

To determine the energy dependent halo profile using these data, we
extracted the halo and PSF radial profiles in energy bands of 200\,eV
width over the range 0.4--10.0\,keV. That is, we use bands that
correspond roughly to the energy resolution of ACIS-I/S, which varies
from 100 to 200\,eV in this range. We find that above 4.0~keV, the
halo intensity is much smaller than the PSF intensity (i.e., the
surface brightness attributable to the point source). Additionally,
the Rayleigh-Gans approximation underestimates the halo intensity
below 1.0\,keV \citep{smith98}. We therefore only use the halo from
1.0\,keV to 4.0\,keV in the following. The total surface brightness
(Halo + PSF), PSF, and PSF-subtracted net halo profiles are shown in
Fig.~\ref{fig:extraction} for the 2--2.2\,keV band. In this band, at
angular distances greater than $10''$, the halo surface brightness is
significantly larger than the surface brightness of the point source
(i.e., the PSF intensity).  On the other hand, for all energies
$>$1.6\,keV and angular distances $<10''$ the data are fully dominated
by the point source and therefore are ignored in the following study.
For energies below 1.6~keV we can approach the point source to
separations as low as $7''$.  At smaller angular separations the
pileup ratio is greater than $\>5\%$ and the data have to be ignored
at all energies.

A final complication arises for energies above 2.0\,keV and radial
distances $>200''$. Here, the halo profile is contaminated by photons
from the $1^\mathrm{st}$ order gratings spectrum. Of special
importance are photons in the MEG$\pm$$1^\mathrm{st}$ order, which
starts closer to the zeroth order than the HEG$\pm$$1^\mathrm{st}$
order spectrum. We therefore ignored the halo radial profile at
$>200''$ for energies above 2.0\,keV.

\section{Spatial Distribution of Dust Grains}
\label{sec:spatial}

\begin{figure}
\plotone{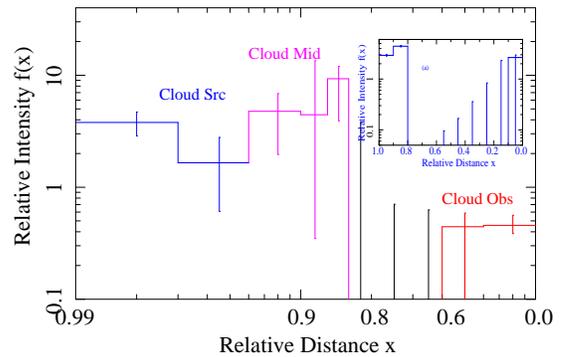}
\caption{Relative density distributions along the LOS based on halo
  radial profile fits using the ZDA BARE-GR-S dust model. The observer
  is at 0, and the source is at 1. A cloud at $\sim$0.88 is
  obvious. The inset (a) figure (in blue) also shows a cloud near the
  source when ten linear-equally spaced bins are instead used in the
  model.  (In subsequent discussion and fits presented below, a
  simplified three-zone model, inspired by the above results, is
  used.) \label{fig:ten}}
\end{figure}

For our initial modeling, dust uniformly distributed along the entire
LOS was assumed. This approach did not result in satisfactory
descriptions of the halo radial profile observations using any of the
18 dust models. Following the method of \citet{xiang05, xiang07}, we
therefore assume an inhomogeneous dust distribution along the line of
sight. We divide the LOS into 10 approximately logarithmically spaced
bins.  That is, the spatial resolution of the dust distribution
increases towards the source. The relative density distribution along
the LOS as derived from fits to the halo radial profile and assuming
the ZDA BARE-GR-S model is shown in Fig.~\ref{fig:ten}.  The ZDA
BARE-GR-S model gave the best fits for the reddening of 4U~1724$-$307
in the work by \citet{valencic09} and it is also one of our favored
models (see Section \ref{sec:discuss}).

The data prefer approximately three regions of different
characteristic densities. The highest density region is found at a
relative distance between $\approx 0.84$--0.94. The regions extending
from $\approx 0.94$--1.0 and from $\approx 0.0--0.6$ exhibited lower
columns. Similar results were obtained when using 10 linearly spaced
bins (inset of Fig.~\ref{fig:ten}): there was little column between
0.0--0.8, and the bins between 0.8--0.9 and 0.9--1.0 have different
characteristic densities.

In the ensuing analysis, we discuss these three characteristic regions
as if they were three individual clouds.  Furthermore, we label these
three regions, starting near us and working outwards towards the
source, as Cloud Obs, Cloud Mid, and Cloud Src.  Cloud Mid contributes
more than 50\% of the total dust grains along the LOS. This result is
consistent with \citet{ling09}, who use a cross-correlation method to
derive the position of this dominant cloud from the delay time of the
scattered photons.

Figure~\ref{fig:spatial} shows that the halo flux between $20''$ and
$45''$ is dominated by photons scattered by Cloud Mid. The delay time
of scattered photons depends especially strongly on the distance of
this cloud. To make the calculation of the predicted distances and
fitting of the halo lightcurves more tractable, we further simplify
the description of the dust spatial distribution. Specifically, based
upon the preliminary radial profile fit shown in Fig.~\ref{fig:ten}
that exhibits three regions with different characteristic densities,
we formally divide the LOS into three components. Uniform density
regions are used to represent Cloud Obs between 0.0--$x_\mathrm{o}$,
Cloud Src between $x_\mathrm{s}$--1.0, and Cloud Mid centered at
$x_\mathrm{c}$. \citet{ling09} claimed limits on the width of this
latter, dominant cloud of $\Delta x=0.016_{-0.006}^{+0.009}$;
therefore, in this work we have adopted a fixed width of 0.016
($\sim$30~pc). We reexamine this limit in \S\ref{sec:distance}, where
we show that a cloud with a large extent does not fit the dust
scattering halo lightcurves.  For convenience, we take the upper limit
to be $x=1$; however, it should be noted that the dust located very
near the source, i.e, $x > 0.99$, might be contaminated by the wind of
HDE~226868, the companion of \cygx1.  Note that the halo scattered by
this dust is concentrated at very small angles ($<$$7''$) and
contributes very little to the portion of the halo considered in this
work. We therefore still use the same dust model along the entire LOS. We
have also checked that changing the upper limit of Cloud Src to $x =
0.99$ from $x=1.0$ does not provide significantly different results
for the $N_{\rm H}$ due to the Cloud Src nor does it affect the
distance determination.

The relative distances $x_\mathrm{o}$, $x_\mathrm{c}$, and
$x_\mathrm{s}$ were initially allowed to vary in our fits.  We found
that both the total hydrogen column density and the position of Cloud
Mid, i.e., $x_\mathrm{c}$, are not greatly affected by variations of
$x_\mathrm{o}$ but are strongly correlated with the value of
$x_\mathrm{s}$.  The former fact is not surprising given the low
relative density in this region close to the observer, as shown in
Fig.~\ref{fig:ten}.  For the following halo radial profile fits we
therefore fix $x_\mathrm{o}$ to 0.70, which is the best fit value for
the model of ZDA BARE-GR-S.  On the other hand, $x_\mathrm{s}$ was
allowed to vary between 0.75--0.98. The eighteen dust grain models
described in Section~\ref{sec:dustmodels} --- MRN, WD01, 15 ZDA models, and
our XLNW model --- are used to fit the halo radial profile.

When the XLNW dust model is used to fit the halo radial profiles, we
derive a scattering hydrogen column density that is consistent with
the absorption hydrogen column density derived from XAFS studies
\citep{lee11}.  It should be noted that our XLNW model yields a
similar position for Cloud Mid as do the BARE-GR-S, and BARE-GR-FG.
These latter models also yield scattering hydrogen column densities
consistent with the XAFS studies.

We now turn to using the above preliminary fits to the halo radial
profile to constrain the fitting of the halo lightcurves.

\section{Distance Determination}
\label{sec:distance}
The halo lightcurves of \cygx1 clearly show significant delays with
respect to the source lightcurves. The delay time strongly depends on
the observed off-axis angles. In order to measure the delay, we
calculate halo light curves for six, $5''$ wide annuli, with the inner
radii ranging from $20$--$45''$. Outside this interval, the delay
times exceed the duration of our observation.

\begin{figure}
\plotone{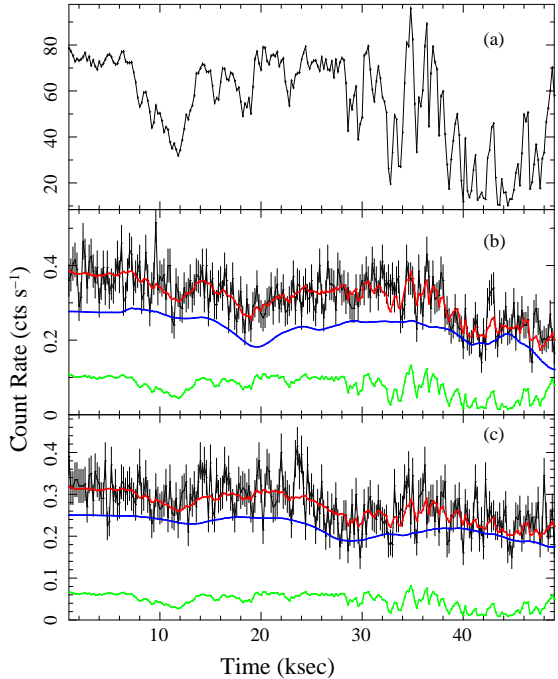}
\caption{Point source lightcurves and lightcurves at a given annulus,
  in the 1--4\,keV band, overplotted by the best-fit models (red)
  using the ZDA BARE-GR-S dust model. The blue line indicates the
  lightcurves derived from the dust model, while the green line is the
  modeled PSF lightcurve.  Their sum yields the red lightcurve.  The
  top panel (a) is the central point source lightcurve.  The middle panel (b)
  is the 20--$25''$ annulus.  The bottom panel (c) is the 30--$35''$
  annulus.  \label{fig:lc}}
\end{figure}

Following the method of \citet{xiang05}, we use the derived dust
spatial distribution to fit the halo lightcurves with the absolute
distance to the source being the free parameter.  The best-fit
distance is determined by minimizing $\chi^2$ based on simultaneously
fitting the delays in \emph{all} lightcurves.  Figure~\ref{fig:lc}
shows the point source and halo lightcurves overplotted with the
best-fit of the ZDA BARE-GR-S model, using a single fit distance for
all lightcurves considered simultaneously.

The time delayed lightcurves also carry information about the
positions of the clouds such that the lightcurve fits also can be used
to constrain the position of $x_\mathrm{s}$
further than by just using the halo profile. We therefore vary $x_\mathrm{s}$
from 0.75 to 0.98, and at each trial value of $x_\mathrm{s}$ we perform a
constrained fit to the radial profile to determine $x_\mathrm{c}$ and the
hydrogen column densities. We then use these radial profile derived
parameters to fit the halo lightcurves and obtain a minimum $\chi^2$.
While the $\chi^2$ from solely a fit of the halo radial profile varies
little with $x_\mathrm{s}$, the $\chi^2$ from fits of the halo lightcurve
varies significantly with $x_\mathrm{s}$. We use this variation in $\chi^2$ to
create a lightcurve determined uncertainty for $x_\mathrm{s}$ (i.e., $\Delta
\chi^2 = 2.71$ to determine the 90\% error bars). For 13 of the 18
dust grain models, the values of $x_\mathrm{s}$ derived from the preliminary
radial profile fits described above and those values derived from the
lightcurves were self-consistent with one another, i.e., the 90\%
error bars for each estimate (lightcurve, radial profile) of $x_\mathrm{s}$
significantly overlapped.

\begin{figure*}
\epsscale{0.8}
\plotone{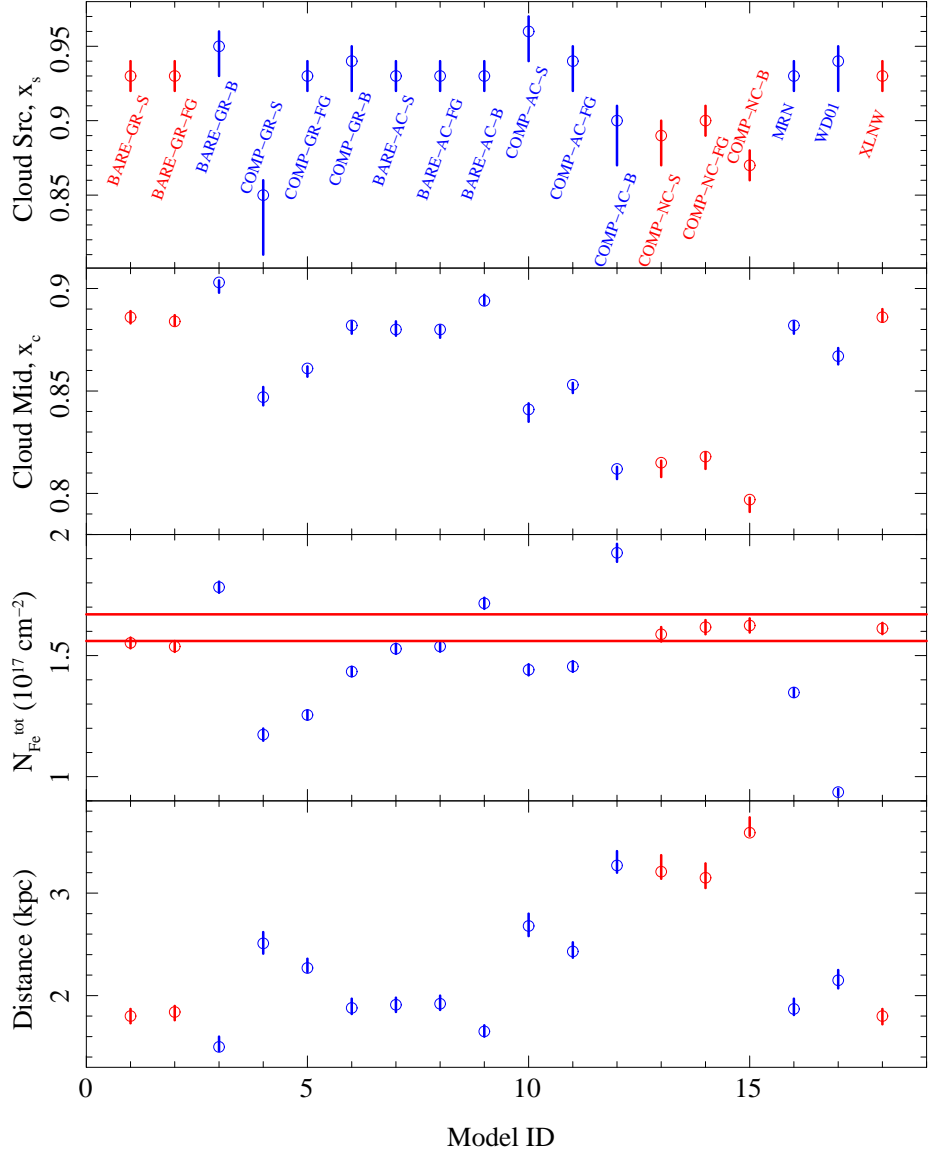}
\caption{The results of the joint fits to the halo radial profile and
  lightcurve for each model.  (Model ID is shown along the $x$-axis,
  and model names are given in the top panel.)  The top two panels
  give the relative positions of the cloud nearest the source
  ($x_\mathrm{s}$), and the next closest cloud ($x_\mathrm{c}$).  The
  next panel gives the summed total neutral column (converted to an
  equivalent Fe column based on the respective olivine abundance
  listed in Table 1 for each dust model) for all of the clouds between
  us and the source.  The two red lines show the iron column estimate
  obtained from the \protect{\cite{lee11}} modeling of the
  $\sim$700~eV FeL absorption edges and associatd XAFS in HETGS
  spectra of \cygx1.  For all panels, models plotted in red agree with
  the neutral column density estimate based on iron. The bottom panel
  gives the best fit source distance for each model.  The median value
  is 1.82\,kpc, while three of the six models that agree with the
  neutral column estimates of \protect{\citet{lee11}} give distance
  estimates that lie in the range from 1.72--1.90\,kpc.
  \label{fig:global_fit}}
\end{figure*}

Given that the iterative approach to fitting the radial profile and
then fitting the lightcurve failed to produce self-consistent fit
parameters for five cases, we then turned to a joint fit of these
data.  Specifically, we used the solutions obtained from the iterative
approach as starting parameter values, and then simultaneously fit
the radial profile and the lightcurve data with six variable
parameters: $x_\mathrm{s}$, $x_\mathrm{c}$, $d$, and three hydrogen column densities,
$N_{\rm H}^{\rm mid}$, $N_{\rm H}^{\rm obs}$, and $N_{\rm H}^{\rm src}$. 
The results of these fits are presented in Table~\ref{tab:nh}. 
The $N_{\rm H}^{\rm tot}$ are the sum of $N_{\rm H}^{\rm mid}$, 
$N_{\rm H}^{\rm obs}$, and $N_{\rm H}^{\rm src}$. The iron
column density $N_{\rm Fe}^{\rm tot}$ is derived from 
$N_{\rm tot}^{\rm H}$ based
on the iron abundance for each dust model (see Table \ref{tab:model}).
Parameter uncertainties are given as the 90\% confidence level for one
interesting parameter, i.e., $\Delta \chi^2 = 2.71$ for fits holding
the parameter of interest frozen at a given value.  Values of $x_\mathrm{s}$,
$x_\mathrm{c}$, $N_{\rm Fe}^{\rm tot}$, and $d$ for each dust model are presented in
Fig.~\ref{fig:global_fit}.

Our derived scattering \emph{dust} iron column density varies from
$0.94\times10^{17}\rm\ cm^{-2}$ to $1.92\times10^{17}\rm\ cm^{-2}$,
which should be equal to the Galactic absorption dust iron column
density. We therefore use results from absorption spectra to constrain
further the dust models. We choose to compare to the iron column
density as it can be measured directly via modeling of the Fe L edge
in high spectral resolution data \citep{manfred09,lee:09b,lee11}. As
discussed by \citet{wilms00}, neutral column absorption in the X-rays
is dominated by $Z>2$ elements, i.e., not hydrogen.  Comparing
directly to the X-ray derived Fe column therefore removes the
ambiguity of the assumed elemental abundances relative to hydrogen in
such spectral models.

The total absorption iron column density of Cyg~X-1 derived from
absorption spectra varies from $0.96\times10^{17}\rm\ cm^{-2}$
\citep{juett06} to $2.52\times10^{17}\rm\ cm^{-2}$
\citep{manfred09}. Aside from possible instrumental differences, these
different values are in part due to variations in the column density
local to the source, e.g., because of absorption in the stellar wind
\citep[see][]{manfred09}, as well as variations in the assumed
absorption cross sections. Our recent studies of observed X-ray
Absorption Fine Structure (XAFS) in the Low/Hard, Transition, and
High/Soft states \citep{lee11} suggest that the dust iron column
density associated with Cyg X-1 is $1.62\pm0.06\times10^{17}\ \rm
cm^{-2}$. Only six of the eighteen dust models considered for the halo
modeling have $N_\mathrm{Fe}$ consistent with this value. (The
parameters for these models are plotted in red in
Fig.~\ref{fig:global_fit}.) Of these six models, three models, ZDA
BARE-GR-S, ZDA BARE-GR-FG, and our proposed model XLNW, yield source
distances that lie in the range from 1.72--1.90\,kpc (The median
distance for all three models is 1.82\,kpc.), whereas three models,
ZDA ZDA COMP-NC-S, ZDA COMP-NC-FG and ZDA COMP-NC-B, yield larger
distances of 3.15--3.59\,kpc.  The large distance values for these
latter three models, compared to the most recent distance estimates
(discussed below) leads us to discount these fits. This leaves us with
three models that we favor to describe the halo and absorption profile
of \cygx1.

\begin{figure}
\plotone{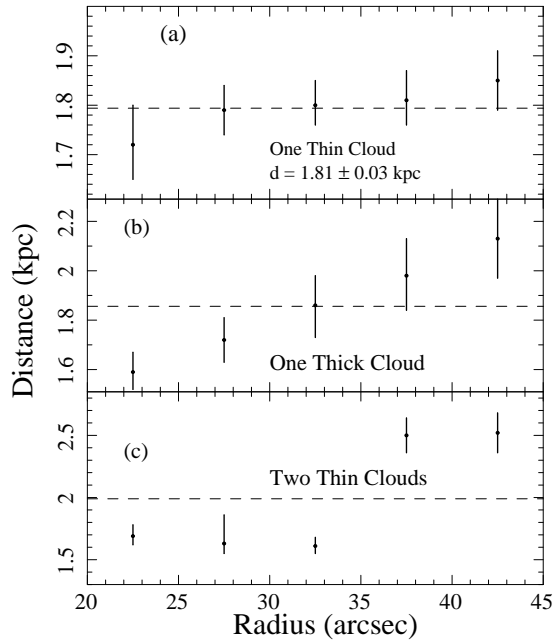}
\caption{The distances derived from lightcurves at different radii for
  one thin cloud (a), one thick cloud (b), and two thin clouds
  (c). Only the distances derived from one thin cloud are consistent
  with each other for each radii. The dashed horizontal lines indicate
  the average distance. The uncertainties in this figure are
  only from the fits to the halo lightcuves. The error bars in this figure
  are only 1$\sigma$. Additional uncertainties
  from uncertainties in the radial profile fits are not
  included.\label{fig:dif}}
\end{figure}

Lastly we consider the possibility that the cloud with a relative
distance between 0.8 and 0.9 is a thick cloud, or that the cloud might
be divided into 2 parts. For the first possibility, we set a cloud
with a width of 0.1 (relative distance) located at $x_\mathrm{c} =
0.878$. This location is the best fit value of $x_\mathrm{c}$ derived
from radial profile fits using the ZDA BARE-GR-S model. The distances
then fitted from halo lightcurves at different angles are then not
consistent with each other, as shown in the middle panel of
Fig. \ref{fig:dif}. For the second possibility, we locate one thin
cloud at 0.85 and another thin cloud at 0.90. The width of each cloud
is set to 0.016. The distances fitted from individual halo lightcurves
are again found to be inconsistent with each other, as shown in the
bottom panel of Fig.  \ref{fig:dif}. We also noticed that the
$\chi^2/$DoF of the distance fit with thick cloud ($\chi^2/$DoF =
1003/777) or two thin clouds ($\chi^2/$DoF = 983/776) is much worse
than that from one thin cloud fit ($\chi^2/$DoF = 884/777). This
further justifies our assumption of a single thin cloud at the
position $x_\mathrm{c}$.

\section{Discussion and Summary}\label{sec:discuss}

Using the X-ray dust scattering halo radial profiles and lightcurves
combined with the X-ray absorption hydrogen column density, we can
begin to distinguish among the dust grain models. Our favored dust
models are BARE-GR-S, BARE-GR-FG, and XLNW. It should be noted that
the first two models, i.e., BARE-GR-S and BARE-GR-FG, also yield the
proper optical extinction for the LOS to the X-ray binary 4U
1724$-$307 \citep{valencic09}. Our fits using the classic MRN model
and the more recent model of WD01 underestimated the hydrogen column
density to \cygx1; similar underestimates with these two models were
found by \citet{valencic08} for X Persei and \citet{valencic09} for 4U
1724$-$307. All three of our favored models contain graphite, while
none of the dust models containing amorphous carbon simultaneously
yield good neutral columns and source distances (see below).  This
result suggests that carbon in the ISM prefers to reside in graphite
instead of amorphous carbon dust. Our proposed model XLNW, which
separates iron from olivine, can yield a proper hydrogen column
density and a proper position for the source, as discussed below. 

Our results show that the dust scattering halo radial profiles
together with the halo lightcurve provide a potentially powerful tool
to determine the relative position of the cloud along the LOS, and to
determine the geometric distance to the point source. All three of our
favored models show that Cloud Src is located at $x_\mathrm{s}
\sim$0.92--0.95 and that Cloud Mid is located at
$x_\mathrm{c}\approx0.885$.  Our fitted distances to \cygx1 for the
most part lie within a narrow range of 1.72--1.90.  This distance
range is consistent with recent radio parallax measurements for \cygx1
which yield a distance of $d=1.86\pm0.12$~kpc \citep{reid2011}. This
is considered the most reliable method to determine the distance. The
distance derived from our proposed dust model containing iron metal
and troilite is consistent with those derived from BARE-GR-S and
BARE-GR-FG, both of which yield the proper column density.

The first estimate of the distance to \cygx1 was made by
\citet{margon73} using the surveyed optical extinction in the field
immediately surrounding \cygx1. The distance of $d=2.5\pm0.4$~kpc was
confirmed by \citet{ninkov87} and was used by many
researchers. \citet{russell07} and \citet{gallo05} imply that
Sharpless 2-101 {hereafter Sh 2-101}, which is is a bright reflection
nebula to the north east of \cygx1, is close to
\cygx1. \citet{russell07} also suggest that the nebula interacts with
the jet of \cygx1. The distance of Sh 2-101 is generally assumed to be
larger than 2.2~kpc. Our results, most of which are significantly
lower than 2.2~kpc, hint that Sh 2-101 is a background source and
therefore the \cygx1 jet is not interacting with it.  Alternatively,
measurements of the distance to Sh 2-101 may be overestimates.

\acknowledgements The authors would like to acknowledge useful
conversations with Randall Smith.  J.C.L gratefully acknowledge
funding support from \chandra\ grant AR8-9007, and the Harvard
University Faculty of Arts and Sciences.  M.A.N was supported by NASA
grants GO8-9036X and SV3-73016.  J.W was partly supported by the
European Commission under contract ITN 215212 ``Black Hole Universe''
and by the Bundesministerium f\"ur Wirtschaft und Technologie through
Deutsches Zentrum f\"ur Luft- und Raumfahrt grants 50OR0701 and
50OR1005.

\bibliographystyle{apj}
\bibliography{xiang}

\begin{thebibliography}{33}
\expandafter\ifx\csname natexlab\endcsname\relax\def\natexlab#1{#1}\fi

\bibitem[\protect\astroncite{{Canizares} et~al.}{2005}]{canizares2005}
{Canizares}, C.~R., et~al., 2005, \pasp, 117, 1144

\bibitem[\protect\astroncite{{Carter} et~al.}{2003}]{carter:2003}
  {Carter}, C., Karovska, M., Jerius, D., Glotfelty, K, \& Beikman,
  S., 2003, ADASS XII, ASP Conference Series Vol. 295,
  eds. H.E. Payne, R.I. JEdzrejewski, \& R.N. Hook, 477.

\bibitem[{{Costantini} {et~al.}(2005){Costantini}, {Freyberg}, \&
  {Predehl}}]{costantini05}
{Costantini}, E., {Freyberg}, M.~J., \& {Predehl}, P. 2005, \aap, 444, 187

\bibitem[{{Gallo} {et~al.}(2005){Gallo}, {Fender}, {Kaiser}, {Russell},
  {Morganti}, {Oosterloo}, \& {Heinz}}]{gallo05}
{Gallo}, E., {Fender}, R., {Kaiser}, C., {Russell}, D., {Morganti}, R.,
  {Oosterloo}, T., \& {Heinz}, S. 2005, \nat, 436, 819

\bibitem[\protect\astroncite{{Garmire} et~al.}{2003}]{garmire:03a}
{Garmire}, G.~P., {Bautz}, M.~W., {Ford}, P.~G., {Nousek}, J.~A., \& {Ricker},
  Jr., G.~R.,  2003,
\newblock in Society of Photo-Optical Instrumentation Engineers (SPIE)
  Conference Series, ed. {J.~E.~Truemper \& H.~D.~Tananbaum}, Vol. 4851, 28

\bibitem[{{Hanke} {et~al.}(2009){Hanke}, {Wilms}, {Nowak}, {Pottschmidt},
  {Schulz}, \& {Lee}}]{manfred09}
{Hanke}, M., {Wilms}, J., {Nowak}, M.~A., {Pottschmidt}, K., {Schulz}, N.~S.,
  \& {Lee}, J.~C. 2009, \apj, 690, 330

\bibitem[{{Henke}(1981)}]{henke81}
{Henke}, B.~L. 1981, in American Institute of Physics Conference Series,
  Vol.~75, Low Energy X-ray Diagnostics, ed. {D.~T.~Attwood \& B.~L.~Henke},
  146--155

\bibitem[{{Juett} {et~al.}(2006){Juett}, {Schulz}, {Charkrabarty},
\& {Gorczyca}}]{juett06} Juett, A.~M., Schulz, 
N.~S., Chakrabarty, D., \& Gorczyca, T.~W.\ 2006, \apj, 648, 1066

\bibitem[\protect\astroncite{{Lee}}{2010}]{lee:10a}
{Lee}, J.~C.,  2010, Space Science Reviews, 157, 93

\bibitem[\protect\astroncite{{Lee} et~al.}{2001}]{lee:01a}
{Lee}, J.~C., {Ogle}, P.~M., {Canizares}, C.~R., {Marshall}, H.~L., {Schulz},
  N.~S., {Morales}, R., {Fabian}, A.~C., \& {Iwasawa}, K.,  2001, Astrophys.
  J., Lett., 554, L13

\bibitem[\protect\astroncite{{Lee} et~al.}{2009a}]{lee:09b}
{Lee}, J.~C., et~al., 2009a,
\newblock in Astro2010: The Astronomy and Astrophysics Decadal Survey, Vol.
  2010,  178

\bibitem[\protect\astroncite{{Lee} et~al.}{2009b}]{lee:09a}
{Lee}, J.~C., {Xiang}, J., {Ravel}, B., {Kortright}, J., \& {Flanagan}, K.,
  2009b, ApJ, 702, 970

\bibitem[{{Lee} {et~al.}(2011)}]{lee11}
{Lee}, J.~C. et al. 2011, in preparation.

\bibitem[{{Lestrade} {et~al.}(1999){Lestrade}, {Preston}, {Jones}, {Phillips},
  {Rogers}, {Titus}, {Rioja}, \& {Gabuzda}}]{lestrade99}
{Lestrade}, J., {Preston}, R.~A., {Jones}, D.~L., {Phillips}, R.~B., {Rogers},
  A.~E.~E., {Titus}, M.~A., {Rioja}, M.~J., \& {Gabuzda}, D.~C. 1999, \aap,
  344, 1014

\bibitem[{{Ling} {et~al.}(2009){Ling}, {Zhang}, {Xiang}, \& {Tang}}]{ling09}
{Ling}, Z., {Zhang}, S.~N., {Xiang}, J., \& {Tang}, S. 2009, \apj, 690, 224

\bibitem[{{Margon} {et~al.}(1973){Margon}, {Bowyer}, \& {Stone}}]{margon73}
{Margon}, B., {Bowyer}, S., \& {Stone}, R.~P.~S. 1973, \apjl, 185, L113

\bibitem[{{Mathis} {et~al.}(1995){Mathis}, {Cohen}, {Finley}, \&
  {Krautter}}]{mathis95}
{Mathis}, J.~S., {Cohen}, D., {Finley}, J.~P., \& {Krautter}, J. 1995, \apj,
  449, 320

\bibitem[{{Mathis} \& {Lee}(1991)}]{mathis91}
{Mathis}, J.~S., \& {Lee}, C. 1991, \apj, 376, 490

\bibitem[{{Mathis} {et~al.}(1977){Mathis}, {Rumpl}, \& {Nordsieck}}]{MRN}
{Mathis}, J.~S., {Rumpl}, W., \& {Nordsieck}, K.~H. 1977, \apj, 217, 425

\bibitem[{{Mauche} \& {Gorenstein}(1984)}]{mauche84}
{Mauche}, C., \& {Gorenstein}, P. 1984, in Bulletin of the American
  Astronomical Society, Vol.~16,  926

\bibitem[{{Ninkov} {et~al.}(1987){Ninkov}, {Walker}, \& {Yang}}]{ninkov87}
{Ninkov}, Z., {Walker}, G.~A.~H., \& {Yang}, S. 1987, \apj, 321, 425

\bibitem[{{Overbeck}(1965)}]{overbeck65}
{Overbeck}, J.~W. 1965, \apj, 141, 864

\bibitem[{{Predehl} {et~al.}(2000){Predehl}, {Burwitz}, {Paerels}, \&
  {Tr{\"u}mper}}]{predehl00}
{Predehl}, P., {Burwitz}, V., {Paerels}, F., \& {Tr{\"u}mper}, J. 2000, \aap,
  357, L25

\bibitem[{{Predehl} \& {Klose}(1996)}]{predehl96}
{Predehl}, P., \& {Klose}, S. 1996, \aap, 306, 283

\bibitem[{{Predehl} \& {Schmitt}(1995)}]{predehl95}
{Predehl}, P., \& {Schmitt}, J.~H.~M.~M. 1995, \aap, 293, 889

\bibitem[{{Reid} {et~al.}(2011){Reid}, {McClintock}, {Narayan}, {Gou},
  {Remillard}, \& {Orosoz}}]{reid2011}
{Reid}, M.~J., {McClintock}, J.~E., {Narayan}, R., {Gou}, L.
             {Remillard}, R.~A. \& Orosz, J.~A.  2011, Science, submitted.

\bibitem[{{Rolf}(1983)}]{rolf83}
{Rolf}, D.~P. 1983, \nat, 302, 46

\bibitem[{{Russell} {et~al.}(2007){Russell}, {Fender}, {Gallo}, \&
  {Kaiser}}]{russell07}
{Russell}, D.~M., {Fender}, R.~P., {Gallo}, E., \& {Kaiser}, C.~R. 2007,
  \mnras, 376, 1341

\bibitem[{{Schulz} {et~al.}(2002){Schulz}, {Cui}, {Canizares}, {Marshall},
  {Lee}, {Miller}, \& {Lewin}}]{schulz02}
{Schulz}, N.~S., {Cui}, W., {Canizares}, C.~R., {Marshall}, H.~L., {Lee},
  J.~C., {Miller}, J.~M., \& {Lewin}, W.~H.~G. 2002, \apj, 565, 1141

\bibitem[{{Smith} \& {Dwek}(1998)}]{smith98}
{Smith}, R.~K., \& {Dwek}, E. 1998, \apj, 503, 831

\bibitem[{{Smith} {et~al.}(2002){Smith}, {Edgar}, \& {Shafer}}]{smith02}
{Smith}, R.~K., {Edgar}, R.~J., \& {Shafer}, R.~A. 2002, \apj, 581, 562

\bibitem[{{Thompson} \& {Rothschild}(2009)}]{thompson09}
{Thompson}, T.~W.~J., \& {Rothschild}, R.~E. 2009, \apj, 691, 1744

\bibitem[{{Tr{\"u}mper} \& {Sch{\"o}nfelder}(1973)}]{trumper73}
{Tr{\"u}mper}, J., \& {Sch{\"o}nfelder}, V. 1973, \aap, 25, 445

\bibitem[{{Valencic} \& {Smith}(2008)}]{valencic08}
{Valencic}, L.~A., \& {Smith}, R.~K. 2008, \apj, 672, 984

\bibitem[{{Valencic} {et~al.}(2009){Valencic}, {Smith}, {Dwek}, {Graessle}, \&
  {Dame}}]{valencic09}
{Valencic}, L.~A., {Smith}, R.~K., {Dwek}, E., {Graessle}, D., \& {Dame}, T.~M.
  2009, \apj, 692, 502

\bibitem[{{Weingartner} \& {Draine}(2001)}]{WD01}
{Weingartner}, J.~C., \& {Draine}, B.~T. 2001, \apj, 548, 296

\bibitem[Wilms et al.(2000)]{wilms00} Wilms, J., Allen, A., 
\& McCray, R.\ 2000, \apj, 542, 914

\bibitem[{{Xiang} {et~al.}(2007){Xiang}, {Lee}, \& {Nowak}}]{xiang07}
{Xiang}, J., {Lee}, J.~C., \& {Nowak}, M.~A. 2007, \apj, 660, 1309

\bibitem[{{Xiang} {et~al.}(2005){Xiang}, {Zhang}, \& {Yao}}]{xiang05}
{Xiang}, J., {Zhang}, S.~N., \& {Yao}, Y. 2005, \apj, 628, 769

\bibitem[{{Zubko} {et~al.}(2004){Zubko}, {Dwek}, \& {Arendt}}]{ZDA}
{Zubko}, V., {Dwek}, E., \& {Arendt}, R.~G. 2004, \apjs, 152, 211

\end{thebibliography}

\begin{deluxetable*}{rlccccccccc}
\tabletypesize{\tiny}
\tablecaption{Distances, cloud positions, and $N_{\rm H}$ from global fits.}
\tablecolumns{11}
\tablehead{\colhead{Model} & 
\colhead{Model} & 
\colhead{Cloud Src$^a$} & 
\colhead{Cloud Mid$^b$} & 
\colhead{Distance} & 
\colhead{$N_{\rm H}^{{\rm mid}}$} & 
\colhead{$N_{\rm H}^{{\rm obs}}$} & 
\colhead{$N_{\rm H}^{{\rm src}}$} & 
\colhead{$N_{\rm H}^{{\rm tot}}$} & 
\colhead{$N_{\rm Fe}^{{\rm tot}}$} & 
\colhead{$\chi^2/DoF$} \\
\colhead{No} & 
\colhead{Name} & 
\colhead{Position ($x_s$)} & 
\colhead{Position ($x_c$)} & 
\colhead{(kpc)}  & 
\multicolumn{4}{c}{$(\rm 10^{21}~cm^{-2})$} & 
\colhead{$(\rm 10^{17}~cm^{-2})$} &
\colhead{} }
\startdata

 1  &  BARE-GR-S  &  $0.93_{-0.01}^{+0.01}$   &  $0.886_{-0.003}^{+0.003}$   &  $1.80_{-0.07}^{+0.07}$   &  $2.47_{-0.04}^{+0.04}$   &   $1.42_{-0.05}^{+0.05}$   &   $0.77_{-0.03}^{+0.03}$   &   $4.66_{-0.07}^{+0.07}$   &   $1.55_{-0.02}^{+0.02}$   &   4277/3969\\
2  &  BARE-GR-FG  &  $0.93_{-0.01}^{+0.01}$   &  $0.884_{-0.002}^{+0.003}$   &  $1.84_{-0.08}^{+0.06}$   &  $2.45_{-0.04}^{+0.04}$   &   $1.40_{-0.04}^{+0.05}$   &   $0.80_{-0.04}^{+0.04}$   &   $4.65_{-0.07}^{+0.07}$   &   $1.54_{-0.02}^{+0.02}$   &   4279/3969\\
3  &  BARE-GR-B  &  $0.95_{-0.02}^{+0.01}$   &  $0.903_{-0.005}^{+0.001}$   &  $1.50_{-0.04}^{+0.10}$   &  $3.32_{-0.04}^{+0.04}$   &   $2.05_{-0.06}^{+0.06}$   &   $0.88_{-0.03}^{+0.03}$   &   $6.25_{-0.08}^{+0.08}$   &   $1.78_{-0.02}^{+0.02}$   &   4291/3969\\
4  &  COMP-GR-S  &  $0.85_{-0.04}^{+0.01}$   &  $0.847_{-0.004}^{+0.005}$   &  $2.51_{-0.10}^{+0.11}$   &  $1.78_{-0.05}^{+0.04}$   &   $1.08_{-0.03}^{+0.04}$   &   $0.70_{-0.05}^{+0.05}$   &   $3.56_{-0.08}^{+0.08}$   &   $1.17_{-0.02}^{+0.03}$   &   4262/3969\\
5  &  COMP-GR-FG  &  $0.93_{-0.01}^{+0.01}$   &  $0.861_{-0.004}^{+0.001}$   &  $2.27_{-0.04}^{+0.09}$   &  $2.18_{-0.03}^{+0.03}$   &   $1.17_{-0.03}^{+0.04}$   &   $0.52_{-0.04}^{+0.04}$   &   $3.87_{-0.06}^{+0.06}$   &   $1.25_{-0.02}^{+0.02}$   &   4253/3969\\
6  &  COMP-GR-B  &  $0.94_{-0.02}^{+0.01}$   &  $0.882_{-0.004}^{+0.002}$   &  $1.88_{-0.06}^{+0.09}$   &  $2.85_{-0.04}^{+0.04}$   &   $1.61_{-0.05}^{+0.05}$   &   $0.70_{-0.04}^{+0.04}$   &   $5.16_{-0.07}^{+0.07}$   &   $1.43_{-0.02}^{+0.02}$   &   4251/3969\\
7  &  BARE-AC-S  &  $0.93_{-0.01}^{+0.01}$   &  $0.880_{-0.003}^{+0.004}$   &  $1.91_{-0.07}^{+0.07}$   &  $2.32_{-0.03}^{+0.03}$   &   $1.47_{-0.04}^{+0.04}$   &   $0.77_{-0.04}^{+0.04}$   &   $4.56_{-0.06}^{+0.06}$   &   $1.53_{-0.02}^{+0.02}$   &   4280/3969\\
8  &  BARE-AC-FG  &  $0.93_{-0.01}^{+0.01}$   &  $0.880_{-0.004}^{+0.002}$   &  $1.92_{-0.06}^{+0.08}$   &  $2.27_{-0.03}^{+0.03}$   &   $1.48_{-0.04}^{+0.04}$   &   $0.75_{-0.04}^{+0.04}$   &   $4.50_{-0.06}^{+0.06}$   &   $1.54_{-0.02}^{+0.02}$   &   4277/3969\\
9  &  BARE-AC-B  &  $0.93_{-0.01}^{+0.01}$   &  $0.894_{-0.002}^{+0.003}$   &  $1.65_{-0.05}^{+0.06}$   &  $2.85_{-0.04}^{+0.04}$   &   $2.09_{-0.06}^{+0.06}$   &   $1.04_{-0.04}^{+0.04}$   &   $5.98_{-0.08}^{+0.08}$   &   $1.72_{-0.02}^{+0.02}$   &   4295/3969\\
10  &  COMP-AC-S  &  $0.96_{-0.02}^{+0.01}$   &  $0.841_{-0.006}^{+0.003}$   &  $2.68_{-0.10}^{+0.12}$   &  $2.61_{-0.03}^{+0.03}$   &   $1.26_{-0.04}^{+0.04}$   &   $0.46_{-0.05}^{+0.05}$   &   $4.33_{-0.07}^{+0.07}$   &   $1.44_{-0.02}^{+0.02}$   &   4263/3969\\
11  &  COMP-AC-FG  &  $0.94_{-0.02}^{+0.01}$   &  $0.853_{-0.004}^{+0.001}$   &  $2.43_{-0.06}^{+0.09}$   &  $2.49_{-0.03}^{+0.03}$   &   $1.32_{-0.04}^{+0.04}$   &   $0.51_{-0.04}^{+0.04}$   &   $4.32_{-0.06}^{+0.06}$   &   $1.45_{-0.02}^{+0.02}$   &   4254/3969\\
12  &  COMP-AC-B  &  $0.90_{-0.03}^{+0.01}$   &  $0.812_{-0.005}^{+0.001}$   &  $3.27_{-0.07}^{+0.14}$   &  $4.43_{-0.07}^{+0.07}$   &   $1.31_{-0.08}^{+0.08}$   &   $1.16_{-0.08}^{+0.08}$   &   $6.90_{-0.13}^{+0.13}$   &   $1.92_{-0.04}^{+0.04}$   &   4289/3969\\
13  &  COMP-NC-S  &  $0.89_{-0.02}^{+0.01}$   &  $0.815_{-0.007}^{+0.001}$   &  $3.21_{-0.07}^{+0.16}$   &  $2.82_{-0.05}^{+0.05}$   &   $1.25_{-0.05}^{+0.05}$   &   $0.68_{-0.06}^{+0.06}$   &   $4.75_{-0.09}^{+0.09}$   &   $1.59_{-0.03}^{+0.03}$   &   4285/3969\\
14  &  COMP-NC-FG  &  $0.90_{-0.01}^{+0.01}$   &  $0.818_{-0.006}^{+0.002}$   &  $3.15_{-0.10}^{+0.14}$   &  $2.84_{-0.05}^{+0.05}$   &   $1.32_{-0.05}^{+0.05}$   &   $0.61_{-0.06}^{+0.05}$   &   $4.77_{-0.09}^{+0.09}$   &   $1.62_{-0.03}^{+0.03}$   &   4282/3969\\
15  &  COMP-NC-B  &  $0.87_{-0.01}^{+0.01}$   &  $0.797_{-0.006}^{+0.001}$   &  $3.59_{-0.03}^{+0.15}$   &  $3.07_{-0.06}^{+0.06}$   &   $1.64_{-0.06}^{+0.06}$   &   $1.06_{-0.06}^{+0.06}$   &   $5.78_{-0.10}^{+0.10}$   &   $1.62_{-0.03}^{+0.03}$   &   4285/3969\\
16  &  MRN  &  $0.93_{-0.01}^{+0.01}$   &  $0.882_{-0.004}^{+0.002}$   &  $1.87_{-0.06}^{+0.10}$   &  $2.07_{-0.03}^{+0.03}$   &   $1.27_{-0.04}^{+0.04}$   &   $0.74_{-0.03}^{+0.03}$   &   $4.08_{-0.06}^{+0.06}$   &   $1.35_{-0.02}^{+0.02}$   &   4303/3969\\
17  &  WD01  &  $0.94_{-0.02}^{+0.01}$   &  $0.867_{-0.004}^{+0.004}$   &  $2.15_{-0.08}^{+0.10}$   &  $1.39_{-0.02}^{+0.02}$   &   $0.84_{-0.02}^{+0.02}$   &   $0.35_{-0.02}^{+0.02}$   &   $2.58_{-0.04}^{+0.04}$   &   $0.94_{-0.01}^{+0.01}$   &   4259/3969\\
18  &  XLNW  &  $0.93_{-0.01}^{+0.01}$   &  $0.886_{-0.002}^{+0.004}$   &  $1.80_{-0.08}^{+0.07}$   &  $2.56_{-0.04}^{+0.04}$   &   $1.47_{-0.05}^{+0.05}$   &   $0.81_{-0.04}^{+0.04}$   &   $4.84_{-0.07}^{+0.07}$   &   $1.61_{-0.02}^{+0.02}$   &   4277/3969\\

 \enddata
\label{tab:nh}
\tablenotetext{a}{Boundary of the region that extends from $x_\mathrm{s}$ to 1, where 0 is the observer position and 1 is the source positon.}
\tablenotetext{b}{Region midpoint. Region extends over $x_\mathrm{c}\pm0.008$.}
\end{deluxetable*}

\end{document}